\documentclass[A4paper,useAMS,usenatbib,fleqn]{mn2e}
\usepackage{graphicx}%for EPS, PS files
\usepackage{amssymb}
\usepackage{times}
\usepackage{enumerate}

\def\aj{AJ}%
\def\apj{ApJ}%
\def\apjl{ApJ}%
\def\apjs{ApJS}%
\def\apss{Ap\&SS}%
\def\aap{A\&A}%
\def\mnras{MNRAS}%
\def\nat{Nature}%
\def\bain{Bull.~Astron.~Inst.~Netherlands}%
\title[On supernova-induced hypervelocity stars]{
Maximum speed of hypervelocity stars ejected from binaries
} 

\author[Tauris]
{Thomas M. Tauris$^{1,2}$\thanks{E-mail: tauris@astro.uni-bonn.de},
\\
$^{1}$ Argelander-Institut f\"ur Astronomie, Universit\"at Bonn, Auf dem H\"ugel 71, D-53121 Bonn, Germany\\
$^{2}$ Max-Planck-Institut f\"ur Radioastronomie, Auf dem H\"ugel 69, D-53121 Bonn, Germany\\
}
\date{Accepted 2014 November 26; Received 2014 September 5}%
\begin{document}

\maketitle

\begin{abstract}
The recent detection of hypervelocity stars (HVSs) as late-type 
B-stars and HVS candidate G/K~dwarfs raises the important question of
their origin. In this Letter, we investigate the maximum possible velocities
of such HVSs if they are produced from binaries which are disrupted via an
asymmetric supernova explosion.
We find that HVSs up to $\sim\!770$ and $\sim\!1280\;{\rm km\,s}^{-1}$
are possible in the Galactic rest frame from this scenario for these two subclasses of HVSs, respectively. We conclude that
whereas a binary origin cannot easily explain all of the observed velocities of
B-type HVSs (in agreement with their proposed central massive black hole origin)
it can indeed account for the far majority (if not all) of the recently detected G/K-dwarf HVS candidates. 
\end{abstract}


%%%%%%%%%%%%%%%%%%%%%%%%%%%%%%%%%%%%%%%%%%%%%%%%%%%%%%%%%%%%%%%%%%%%%%%%%%%%%%%%

\section{Introduction}\label{sec:intro}
\begin{figure}
\centering
\includegraphics[width=0.47\columnwidth,angle=-90]{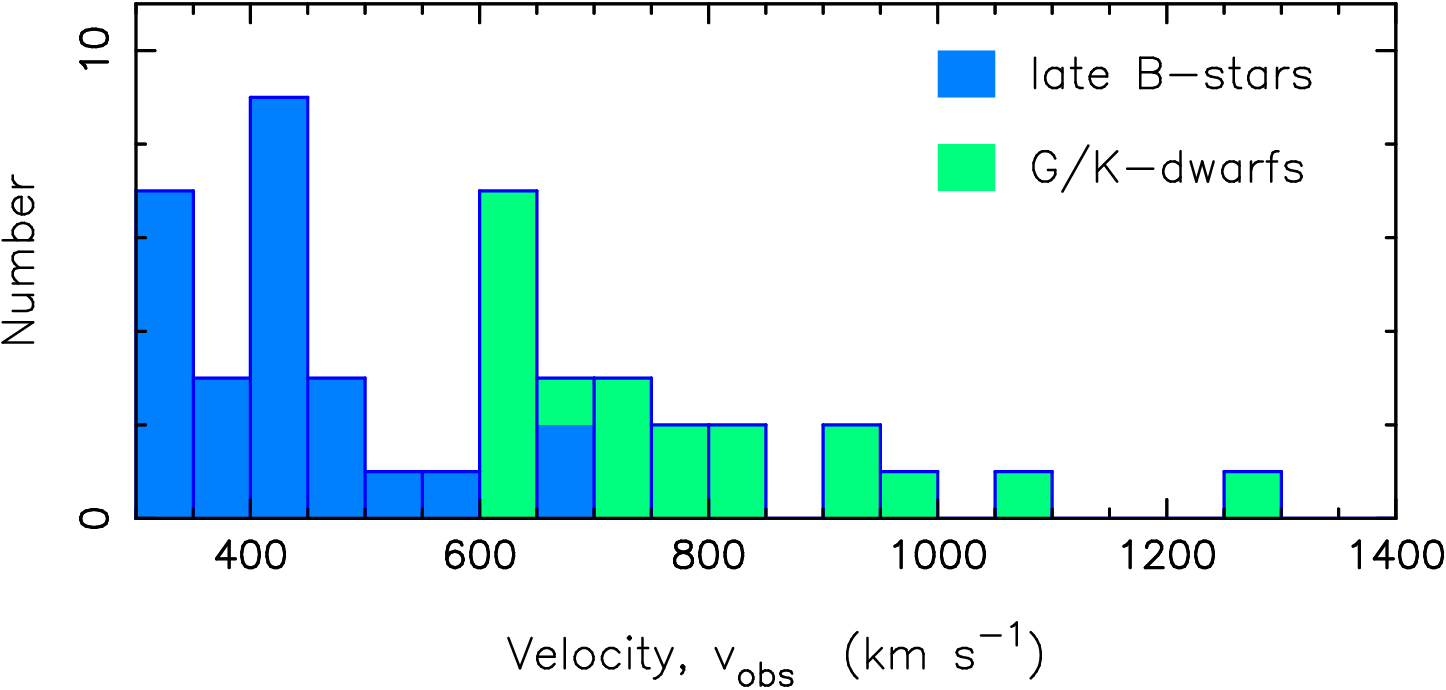}
\caption{Velocity distribution of late B-type HVSs \citep{bgk14} and G/K-dwarf HVS candidates \citep{psh+14}. 
         The B-star velocities are 
         only based on 1D radial velocities (with typical uncertainties of $\sigma = \pm 10\;{\rm km\,s}^{-1}$). 
         The G/K~dwarfs are selected on high proper-motion measurements ($\sigma = \pm 100\;{\rm km\,s}^{-1}$) and may suffer from distance errors.  
         All velocities are given with respect to the Galactic rest frame.
         The Galactic escape velocity is about $350-400\;{\rm km\,s}^{-1}$ in the halo \citep{kbgb08} and
         $\sim\!550\;{\rm km\,s}^{-1}$ in the local disc \citep{srh+07} -- appropriate for many of the observed B-type and G/K-dwarf HVSs, respectively.
  }
\label{fig:obs}
\end{figure}

In recent years, a large number of hypervelocity star (HVS) candidates have been 
reported \citep[e.g.][and references therein]{bgkk05,bgkk06,enh+05,hhob05,bgk09,bgk12,bgk14,tpsh09,llz+12,psh+14,zcl+14}. 
Here we define genuine HVSs only as stars which will escape the gravitational potential of our Galaxy.
Depending on the location and direction of motion, this criterion typically corresponds to a
stellar velocity in the Galactic rest frame $>\!400\;{\rm km\,s}^{-1}$ \citep{kbgb08}.
More than 50 stars can thus be classified as HVSs  
-- see Fig.~\ref{fig:obs} for a subsample.
 
HVSs can obtain their large velocities\footnote{Here we do not distinguish velocity from speed (magnitude of velocity).} 
from a number of different processes.
\citet{hil88} predicted the formation of HVSs via tidal disruption
of tight binary stars by the central supermassive black hole (SMBH) of the Milky Way.
In this process one star is captured by the SMBH while the other is ejected at high speed
via the gravitational slingshot mechanism.
Also exchange encounters in other dense stellar environments \citep[e.g.][]{aar74} between hard binaries and 
massive stars may cause stars to be ejected and escape our Galaxy \citep{leo91,ggp09}.
A competing mechanism for producing HVSs is disruption
of close binaries via supernova (SN) explosions \citep{bla61,boe61,tt98,zwg13}. As demonstrated
in \citet{tt98}, the runaway velocities of both ejected stars can reach large values when 
asymmetric SNe are considered, i.e. when the newborn neutron star (NS) receives a momentum kick at birth. 

The nature of the HVSs spans a wide range of types from OB-stars, to metal-poor F-stars and G/K~dwarfs.
While there is evidence from many late-type~B HVSs in the halo to originate from the Galactic SMBH \citep[e.g.][]{bgk14}
other HVSs seem to originate from the Galactic disc \citep[e.g.][]{hen+08,llz+12,psh+14}.
This calls for a detailed analysis to explain the kinematic origin of, in particular, the latter group.

In this Letter, we investigate the maximum possible ejection velocities of HVSs with different masses originating
from disrupted binaries via asymmetric core-collapse SNe. We use Monte Carlo techniques and perform a systematic investigation
of the parameter space prior to/during the SN in order to probe the resulting velocities. 
The effects of SN shell impact on the companion star are included in our calculations. 
A particular focus is given to late-type B and G/K-dwarf HVSs. 
In Section \ref{sec:SN_model} we briefly describe our model. Our results are presented in Section~\ref{sec:results} and a 
discussion follows in Section~\ref{sec:discussions}. Our conclusions are summarized in Section~\ref{sec:summary}.

%%%%%%%%%%%%%%%%%%%%%%%%%%%%%%%%%%%%%%%%%%%%%%%%%%%%%%%%%%%%%%%%%%%%%%%%%%%%%%%%

\section{Modelling the dynamical effects of SNe}\label{sec:SN_model}
\citet{tt98} derived analytical formulae to calculate the velocities of stars ejected
from binaries in which asymmetric SNe occur. 
The velocity of the ejected companion star, $v_2$ depends on: 
 the pre-SN orbital separation, $r$; 
 its mass, $M_2$; 
 the mass of the exploding star, $M_{\rm He}$ (in close binaries often a stripped helium star with a mass of $3-5\;M_{\odot}$); 
 the mass of the stellar compact remnant (for a NS, typically $M_{\rm NS}\simeq 1.4\;M_{\odot}$); 
 the mass and the velocity of the ejected SN shell, $M_{\rm ejecta}$ and $v_{\rm ejecta}$, 
 and the resulting impact velocity on the companion star caused by the ejected shell, $v_{\rm im}$ (which depends on the explosion energy, $E_{\rm SN}$);
 and finally, the kick velocity (magnitude and direction) imparted on the newborn NS, $\vec{w}$. 

The major component affecting the maximum possible ejection velocity of the companion star, $v_2^{\rm max}$ is its pre-SN orbital velocity, $v_{\rm 2,orb}$.
Hence, to produce ejected HVSs it is clear that very close binaries are needed.

The minimum separation between the two stars prior to the SN explosion is assumed to be limited by the Roche-lobe radius of
the companion star. The reason for this is that the exploding star itself often fills its 
Roche lobe \citep[via so-called Case~BB Roche-lobe overflow, see e.g.][and references therein]{tv06}.  
Thus to avoid the onset of a common envelope \citep[e.g.][]{ijc+13}, which may lead to a fatal merger event prior to the explosion, 
we must require that the companion star does not fill its Roche lobe too. 
Another required constraint in this investigation is that the post-SN trajectory of the NS will not lead to a merger event
in an otherwise disrupted system, i.e. if the periastron separation, $q$ in case of motion along an inbound leg of 
the hyperbolic orbit \citep[$\gamma >0$ and $\xi > 2$, in the notation of][]{tt98} 
is smaller than the radius of the companion star, $R_2$ then the system will likely merge and not produce a HVS.

We applied equations~(51--56) of \citet{tt98} on a large range of binary systems to systematically investigate the maximum possible ejection velocities, $v_2^{\rm max}$ 
by using Monte Carlo simulations. 
The effects of the SN shell impact are discussed in Appendix~A, and a quick sanity check on the applied equations is given in Appendix~B.

%%%%%%%%%%%%%%%%%%%%%%%%%%%%%%%%%%%%%%%%%%%%%%%%%%%%%%%%%%%%%%%%%%%%%%%%%%%%%%%

\section{Results}\label{sec:results}
\begin{figure}
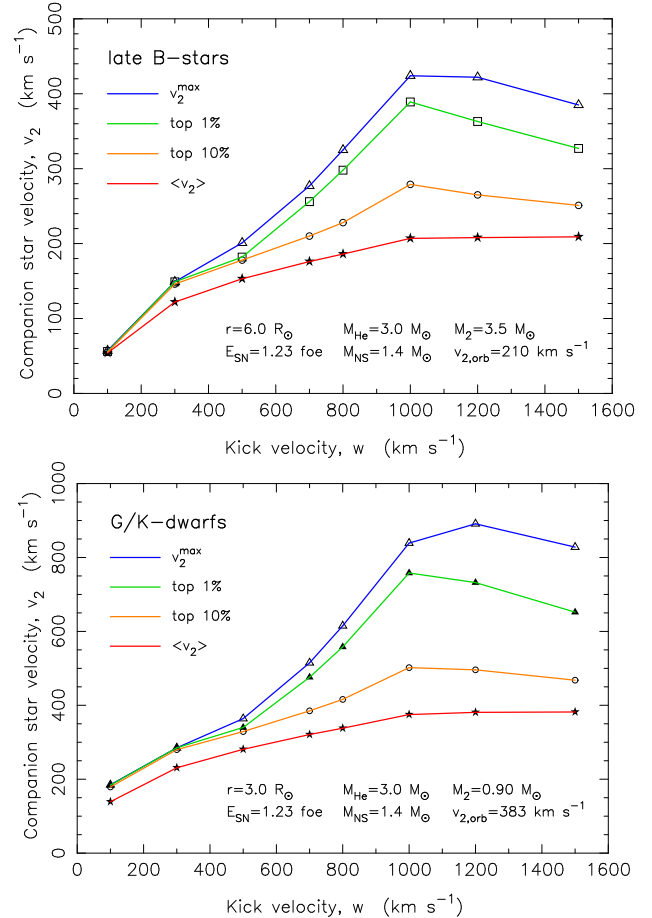

\centering
\includegraphics[width=0.72\columnwidth,angle=-90]{v_HVS1.ps}\\
\includegraphics[width=0.72\columnwidth,angle=-90]{v_HVS2.ps}
\caption{Simulated ejection velocities of B-star companions ($3.5\;M_{\odot}$, top) and G/K-dwarf star companions ($0.9\;M_{\odot}$, bottom)
         as a function of the kick velocity magnitude, $w$ imparted on the NS.
         In both cases we assumed $M_{\rm He}=3.0\;M_{\odot}$, $M_{\rm NS}=1.4\;M_{\odot}$ and $E_{\rm SN}=1.23\times 10^{51}\;{\rm ergs}$
         [note, 1~foe = $10^{51}\;{\rm ergs}$]. 
         The initial (pre-SN) orbital separations were 6.0 and $3.0\;R_{\odot}$, respectively.
         The four curves in each panel represent the maximum velocities obtained, $v_2^{\rm max}$ (blue), the top~1~per~cent (green), the top~10~per~cent (orange)
         and the average values, $\langle v_2 \rangle$ (red), respectively. 
  }
\label{fig:v2}
\end{figure}

In Fig.~\ref{fig:v2}, we have plotted the resulting values of $v_2$ as a function of kick velocity magnitudes, $w$
imparted on the NS for one specific set of initial parameters applied to late-type B-star companions
with an initial mass of $3.5\;M_{\odot}$ (top) and G/K-dwarf companions with an initial mass of $0.9\;M_{\odot}$ (bottom). 
In both cases we simulated $10^6$ explosions using an isotropic kick distribution.
All ejection velocities in this Letter ($v_2$ and $v_{\rm NS}$) are quoted with respect to the c.m. reference frame of the pre-SN binary.
It is interesting to notice that the values of $v_2^{\rm max}$ peak when $w\simeq 1000-1200\;{\rm km\,s}^{-1}$ whereas 
the average values, $\langle v_2 \rangle$ keep increasing with $w$ (approaching an asymptotic value 
as $w\rightarrow \infty$, cf.~Appendix~B).  

It is also seen that the values of $v_2$ for G/K-dwarf stars are generally twice as large as those for the late-type B-stars. 
This is mainly caused by the difference in the applied pre-SN orbital separations, $r$. The G/K~dwarfs are able to remain
much closer to the exploding star without filling their Roche lobes ($r_{\rm min}\simeq 2.9\;R_{\odot}$, for an exploding
star mass of $M_{\rm He}=3.0\;M_{\odot}$) compared to the more massive, 
and larger, B-stars ($r_{\rm min}\simeq 5.6\;R_{\odot}$, for $M_{\rm He}=3.0\;M_{\odot}$).
Furthermore, the G/K~dwarfs have smaller masses, $M_2$.
As a result of these effects, the G/K~dwarfs have larger orbital velocities, $v_{\rm 2,orb}$ prior to the SN,
explaining their larger values of $v_2$.

\begin{figure}
\centering
\includegraphics[width=0.72\columnwidth,angle=-90]{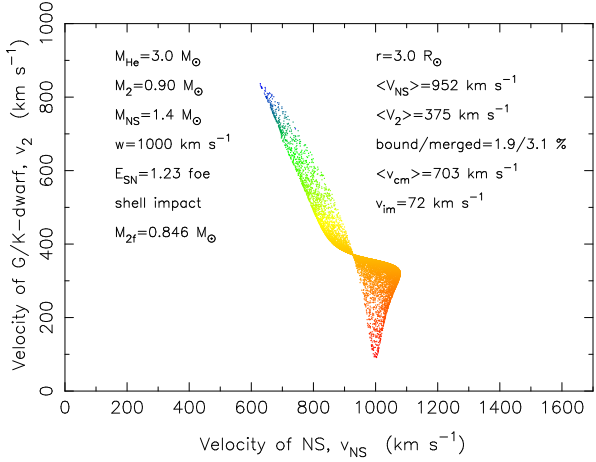}
\caption{Ejection velocities of G/K-dwarf companions, $v_2$ plotted as a function of the ejection velocities of the newborn NSs, $v_{\rm NS}$.
         The plot shows $10^4$~SNe whereas the legend numbers are based on $10^6$~SNe.
         The plot was calculated for $M_2=0.9\;M_{\odot}$, $w=1000\;{\rm km\,s}^{-1}$,
         $M_{\rm He}=3.0\;M_{\odot}$, $M_{\rm NS}=1.4\;M_{\odot}$, $E_{\rm SN}=1.23\times 10^{51}\;{\rm ergs}$ and $r=3.0\;R_{\odot}$.
  }
\label{fig:vns-v2}
\end{figure}

In Fig.~\ref{fig:vns-v2}, we plot the distribution of $v_2$ and $v_{\rm NS}$ for the systems with ejected G/K-dwarf stars
plotted in the bottom panel of Fig.~\ref{fig:v2}, and for which $w=1000\;{\rm km\,s}^{-1}$. 
In this case, as a result of the shell impact, the companion star decreases its mass from 0.90 to $0.846\;M_{\odot}$.
About 1.9~per~cent of the systems survived as bound binaries, with an average systemic velocity of an impressive $703\;{\rm km\,s}^{-1}$,
and 3.1~per~cent of the systems merged as a result of the SN.
The average velocity of the ejected G/K~dwarfs is $\langle v_2 \rangle = 375\;{\rm km\,s}^{-1}$. However, the entire interval of
possible values of $v_2$ spans between 87 and $839\;{\rm km\,s}^{-1}$ for this particular setup.

\subsection{Dependence on kick direction}
\begin{figure}
\centering
\includegraphics[width=0.98\columnwidth,angle=0]{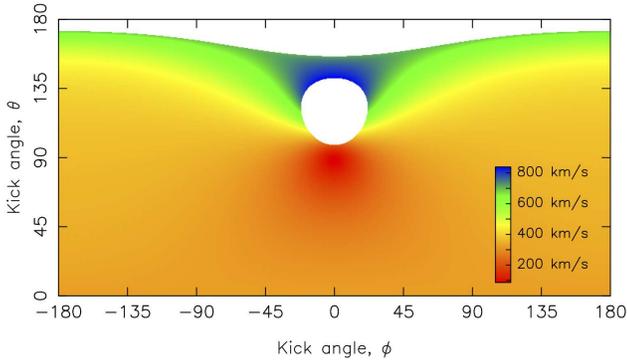}
\caption{Ejection velocities of G/K-dwarf companions plotted as a function of the kick direction angles, $\theta$ and $\phi$.
         The colours represent the resulting ejection velocities, $v_2$ varying between $87\;{\rm km\,s}^{-1}$ (red)
         and $839\;{\rm km\,s}^{-1}$ (blue).
         The plot was calculated for the same initial parameters as in Fig.~\ref{fig:vns-v2}.
  }
\label{fig:angles}
\end{figure}

In Fig.~\ref{fig:angles}, we show that the value of $v_2$ is highly dependent on the direction of the kick imparted on the NS.
The white area in the middle corresponds to cases where the newborn NS is shot into the companion star and the system is assumed to merge.
For highly retrograde kick directions ($\theta \rightarrow 180^\circ$) the systems remain bound, hence the white area at the top of the plot
(the shape of which depends on $r$ and $w$).
The SN-induced HVSs with the largest values of $v_2$ are ejected close to the plane of the pre-SN binary ($\phi=0$),
causing their spin axis to be almost perpendicular to their velocity vector.
While the kick angle $\phi$ is chosen randomly, the polar kick angle, $\theta$ is weighted by $\sin \theta$
in order to obtain an isotropic kick direction \citep[see fig.~1 in][]{tt98}.

\subsection{Dependence on pre-SN parameters}
\begin{figure}
\centering
\includegraphics[width=0.72\columnwidth,angle=-90]{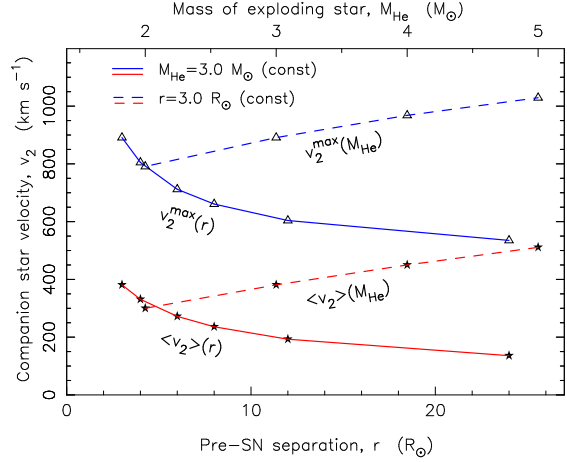}\\
\caption{Simulated ejection velocities of G/K-dwarf star companions ($M_2=0.9\;M_{\odot}$)
         as a function of the pre-SN orbital separation, $r$ (solid lines) and mass of the exploding star, $M_{\rm He}$ (dashed lines). 
         Here we assumed $w=1200\;{\rm km\,s}^{-1}$, $E_{\rm SN}=1.23\times 10^{51}\;{\rm ergs}$ and $M_{\rm NS}=1.4\;M_{\odot}$. 
         The blue and red lines represent $v_2^{\rm max}$ and $\langle v_2 \rangle$, respectively.
  }
\label{fig:dependence}
\end{figure}

As already hinted, besides the kick $\vec{w}$, the value of $v_2$ also depends on pre-SN parameters, in particular the pre-SN
orbital separation, $r$ and the mass of the exploding star, $M_{\rm He}$ (see also Eq.~\ref{eq:v2proxy}). This is demonstrated in Fig.~\ref{fig:dependence}.
The value of $v_2$ decreases both with increasing values of $r$ and decreasing values of $M_{\rm He}$. 
On the other hand, the dependence of $v_2$ on the explosion energy, $E_{\rm SN}$ is quite weak.
For example, using a constant $w=1200\;{\rm km\,s}^{-1}$ 
and increasing the value of $E_{\rm SN}$ from 1.23 to 8~foe, only causes $v_2^{\rm max}$ to increase by $\sim\!6$~per~cent from 891 to $943\;{\rm km\,s}^{-1}$.
The reason for this is that $v_{\rm 2,orb}^2 \gg v_{\rm im}^2$ despite the increase in $E_{\rm SN}$.

We emphasize that our aim here is solely to calculate $v_2^{\rm max}$. The general values of $v_2$ are expected to be much smaller.
Although a detailed population synthesis is beyond the scope of this Letter, we made a simple test and
found $\langle v_2 \rangle = 168\;{\rm km\,s}^{-1}$, if we choose $r$ randomly between 3 and $30\;R_{\odot}$, and only apply $w=450\;{\rm km\,s}^{-1}$.

%%%%%%%%%%%%%%%%%%%%%%%%%%%%%%%%%%%%%%%%%%%%%%%%%%%%%%%%%%%%%%%%%%%%%%%%%%%%%%%

\section{Discussions}\label{sec:discussions}
We have investigated the maximum runaway velocities of SN-induced HVSs. 
However, one must bear in mind to add the Galactic rotational velocity (typically of the order of $v_{\rm Gal}^{\rm rot}\simeq \pm\,230\;{\rm km\,s}^{-1}$) 
at the birth location of the binary system. Hence, in the Galactic rest frame we obtain
$v_{2, {\rm grf}}^{\rm max}\simeq v_2^{\rm max} + 230\;{\rm km}\,{\rm s}^{-1}$.

\subsection{B-type HVSs}
From our simulations we find that only under the most extreme favourable conditions (with respect to $r$, $w$, $\theta$, $\phi$, $M_{\rm He}$
and $E_{\rm SN}$) is it possible for a late-type B-star ($\sim\!3.5\;M_{\odot}$) to achieve $v_2^{\rm max}$ up to $\sim\!540\;{\rm km\,s}^{-1}$
(in those particular cases $v_{\rm im}=110\;{\rm km\,s}^{-1}$ and the final post-ablation stellar mass is $\sim\!3.24\;M_{\odot}$).
This value implies that $v_{2, {\rm grf}}^{\rm max}\sim 770\;{\rm km}\,{\rm s}^{-1}$.
Therefore, any observed B-type HVS which does not exceed this velocity at its origin, 
and whose trajectory does not point to the central SMBH, could potentially be the result of a disrupted binary.
However, we caution that the effect of adding $v_{\rm Gal}^{\rm rot}$ is less important for the HVSs high in the halo and also
note that these stars lose some of their kinetic energy along the trajectory to their current location.

Finally, we have investigated disrupted binaries with early, massive B-star companions ($10\;M_{\odot}$). 
Here we find $v_2^{\rm max}\sim\! 320\;{\rm km\,s}^{-1}$, corresponding to $v_{2, {\rm grf}}^{\rm max}\sim 550\;{\rm km}\,{\rm s}^{-1}$, 
a significantly lower value compared to the late-type B-stars. 

\subsection{G/K-dwarf HVS candidates}
We can reproduce G/K-dwarf HVSs with runaway velocities up to $v_2^{\rm max}\simeq 1050\;{\rm km\,s}^{-1}$ 
(in those extreme cases $v_{\rm im}=195\;{\rm km\,s}^{-1}$, and the final post-ablation stellar mass is $\sim\!0.71\;M_{\odot}$
for a pre-SN mass of $0.90\;M_{\odot}$).
Hence, in the Galactic rest frame $v_{2, {\rm grf}}^{\rm max}\sim 1280\;{\rm km}\,{\rm s}^{-1}$. 
Such high velocities can certainly explain many, if not all, of the recently discovered G/K-dwarf HVS candidates \citep{psh+14}. Interestingly enough,
these HVSs do not seem to originate from the centre of our Milky~Way, bringing further support for a disrupted binary scenario as to their origin.

\subsection{Kick velocities of newborn NSs}
In this work the aim has been to calculate the maximum possible runaway velocities for HVSs ejected from disrupted binaries via 
asymmetric SNe. The magnitude of the kick, $w$ has been treated as a free parameter (besides the assumption of isotropy in the kick direction) 
and we find peak values of $v_2^{\rm max}$ for $w=1000-1200\;{\rm km\,s}^{-1}$. An important question, however, is if such large kicks are realistic?
Although the average kick velocities seem to be of the order $400-500\;{\rm km\,s}^{-1}$ \citep[inferred from studies of proper motions of
young radio pulsars,][]{ll94,hllk05} there are NSs which have received significantly larger kicks.
These include the radio pulsars B2011+38 and B2224+65 which (depending on their precise distances) both have 2D velocities exceeding $1500\;{\rm km\,s}^{-1}$ \citep{hllk05}.
The latter pulsar is observed with a bow shock (the ``guitar nebula'') which confirms that it is moving with a large velocity \citep{crl93}.
Another supersonic runaway pulsar with a velocity in excess of $1000\;{\rm km\,s}^{-1}$ is IGR~J11014$-$6103 \citep{tbr+12,pbp+14}.
Finally, B1508+55 has an fairly precise measured velocity of $\sim\!1100\pm100\;{\rm km\,s}^{-1}$ based on VLBA measurements of its proper motion and parallax \citep{cvb+05}.

Further evidence for large kicks can be found from combining simulations of the dynamical effects of SNe with
future observations of X-ray binaries with large systemic velocities, following the recipe outlined by \citet{tfv+99}. 
Given the above-mentioned evidence for large kicks we therefore predict the existence of low-mass X-ray binaries (and binary millisecond pulsars) 
with peculiar systemic velocities in excess of $700\;{\rm km\,s}^{-1}$, cf. Section~\ref{sec:results}.

Theoretical simulations of SNe \citep[see][for a review]{jan12} can also account for kicks in excess of $1000\;{\rm km\,s}^{-1}$ \citep[e.g.][]{skjm06,wjm13}.
Hence, our HVS simulations presented in this Letter are based on a solid foundation of evidence for the possibility of large kicks.

%%%%%%%%%%%%%%%%%%%%%%%%%%%%%%%%%%%%%%%%%%%%%%%%%%%%%%%%%%%%%%%%%%%%%%%%%%%%%%%

\section{Summary}\label{sec:summary}
We have performed systematic Monte Carlo simulations to investigate the maximum possible runaway velocities of HVSs ejected from
disrupted binaries via asymmetric SNe. For companion stars with initial masses of 0.90, 3.50 and $10\;M_{\odot}$
we find $v_{2, {\rm grf}}^{\rm max}=1280$, 770 and $550\;{\rm km\,s}^{-1}$, respectively. 
While a significant fraction of late B-type HVS have been shown in the literature to originate from the central SMBH \citep{bgk14}, we have presented evidence that,
in particular, the majority (if not all) of the presently observed G/K-dwarf HVS candidates could very well originate from a binary disruption scenario.
However, we caution that a more robust conclusion on the rates and the distribution of $v_{2, {\rm grf}}$ requires detailed population synthesis studies. 

Finally, we note that a firm identification of a HVS being ejected from a binary via a SN is still missing, although
a candidate (HD~271791) has been proposed by \citet[]{pnhb08}; however, see also the interpretation of \citet{gva09}.

\section*{Acknowledgements}
TMT cordially thanks Warren Brown and Pablo Marchant for comments and also acknowledges 
the receipt of DFG grant TA~964/1-1.

%%%%%%%%%%%%%%%%%%%%%%%%%%%%%%%%%%%%%%%%%%%%%%%%%%%%%%%%%%%%%%%%%%%%%%%%%%%%%%%%
%\bibliographystyle{mn2e}
%\bibliography{tauris_refs}
\footnotesize{

}
\normalsize

%%%%%%%%%%%%%%%%%%%%%%%%%%%%%%%%%%%%%%%%%%%%%%%%%%%%%%%%%%%%%%%%%%%%%%%%%%%%%%%%
\newpage
\appendix
\section{The supernova shell impact}
The formulae of \citet{tt98} include the impact of the SN shell on the companion star. This effect is significant and must be included when
probing the maximum possible ejection velocities since these arise from the
tightest pre-SN systems in which the companion star is relatively close to
the exploding star (and hence the cross-section for absorbing momentum of the SN ejecta
is relatively high). 
To model the SN shell impact we adopt a modified version of the analytical formulae of \citet{wlm75}, 
following the implementation in \citet{tt98}. 
While there are several studies published on the shell impact on the companion star in single-degenerate Type~Ia SNe,
no systematic studies are yet available on similar effects caused by core-collapse SNe (Type~Ib/c SNe)\footnote{Although we are aware
of work in progress (Z.-W. Liu et al.). Recently, \citet{hsy14} presented a novel case with core-collapse SN ejecta interacting with
a $10\;M_{\odot}$ giant star companion. However, such a wide orbit (and such a massive companion star) is not suitable for our purposes here.}. 
Therefore, we use a new calibration based on more recent multidimensional hydrodynamical simulations
of the impact of the ejected shell in single-degenerate Type~Ia SNe systems \citep[cf.][]{mbf00,prwh08,prt12,lpr+12}.

\begin{figure}
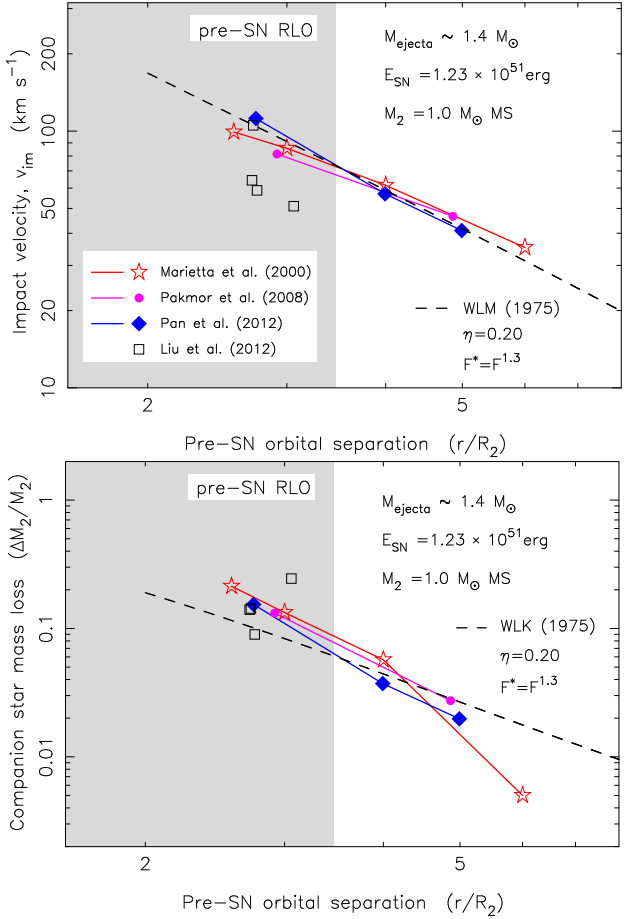

\centering
\includegraphics[width=0.72\columnwidth,angle=-90]{vim_calibration_paper.ps}\\
\includegraphics[width=0.72\columnwidth,angle=-90]{Mstrip_calibration_paper.ps}
\caption{Estimates of impact velocity, $v_{\rm im}$ (top) and fractional mass loss from the companion star, $F=\Delta\,M_2/M_2$ (bottom)
         as a function of pre-SN orbital separation in units of the companion star radius $(r/R_2)$.
         Our analytical formulae (dashed lines) result from a modified prescription of \citet{wlm75} calibrated on a $1.0\;M_{\odot}$ main sequence star companion. 
         The various symbols show results based on hydrodynamical simulations of $0.74-1.22\;M_{\odot}$ companion stars. 
         These calculations were based on Type~Ia SNe and thus the ejecta mass, $M_{\rm ejecta}\simeq 1.4\;M_{\odot}$ (total disruption
         of a Chandrasekhar-mass white dwarf). The explosion energy is in all cases assumed to be $1.23\times 10^{51}\;{\rm ergs}$ \citep[model W7 of][]{nty84}. 
         The shell effects are strongly decreasing with increasing pre-SN orbital separation.
         The grey-shaded zone corresponds to systems where a $1.0\;M_{\odot}$ companion star (the pre-SN donor star) overfills its Roche lobe
         in a binary with a $1.4\;M_{\odot}$ accretor (as in a pre-SN~Ia binary).
  }
\label{fig:SN-strip}
\end{figure}
In Figure~\ref{fig:SN-strip} we show our calibrated fits (dashed lines) given by a slight rewriting of \citet{wlm75}:
\begin{eqnarray}\label{eq:vim} 
  v_{\rm im} & = & 0.20\,v_{\rm ejecta}\,\left( R_2/2r\right)^2\;\left(M_{\rm ejecta}/M_2\right)\,x_{\rm crit}^2 \times \\ \nonumber
             &   & \frac{1+\ln (2\,v_{\rm ejecta}/v_{\rm esc})}{1-F^\ast} ,
\end{eqnarray}
where $R_2$ is the radius of the companion star, $v_{\rm esc}=\sqrt{2GM_2(x_{\rm crit})/R\,x_{\rm crit}}$ is the surface escape
velocity of the companion star (typically, $800-1\,000\;{\rm km\,s}^{-1}$), and the parameter $x_{\rm crit}$
is a critical fraction of the radius outside of which a total mass fraction, $F_{\rm strip}$ is stripped and inside
of which a certain fraction, $F_{\rm ablated}$ is ablated. 
The latter parameters are found from fitted functions to the tabulated values of \citet{wlm75}.
After the SN shell impact, the new mass of the companion star is given by: $M_{2f}=M_2\,(1-F^\ast)$, where
we have adopted $F^\ast=(F_{\rm strip}+F_{\rm ablate})^{1.3}$.
Finally, the average ejecta velocity, $v_{\rm ejecta}$ (typically, $8\,000-10\,000\;{\rm km\,s}^{-1}$) is simply estimated from
$(2\,E_{\rm SN}/M_{\rm ejecta})^{0.5}$, where $E_{\rm SN}$ is the explosion energy of the SN \citep[although this expression is 
probably an overestimate by about 10~per~cent compared to the results from hydrodynamical studies of SNe, cf.][]{mbf00}.

Strictly speaking, $v_{\rm im}$ is an {\it effective} velocity assuming an instant addition of
incident shell momentum (stripping) and subsequent momentum resulting from mass loss due to ablation 
of stellar material from the surface layers heated by the passing shock wave. 
Although in a close binary the incident energy of the SN debris may exceed the binding energy of the companion star by a few orders of magnitude, 
this energy is deposited in the outer layers of the star whereby a main-sequence companion star can easily survive such
an impact \citep{che74,wlm75}.

\section{A sanity check on the equations}
A quick sanity check on our applied equations~(51--56) of \citet{tt98} is made via their equations~(13) and (44--47), and 
reveals the results for the limiting cases with either no remnant mass (as in the case of SNe~Ia) or an infinite high kick velocity: 
\begin{equation}
  \lim_{M_{\rm NS} \to 0} v_2 = \sqrt{v_{\rm 2,orb}^2 + v_{\rm im}^2} ,
\label{eq:M20}
\end{equation}
\begin{equation}
  \lim_{w \to \infty} v_2 = \sqrt{v_{\rm 2,orb}^2 + v_{\rm im}^2} \qquad \wedge \qquad \lim_{w \to \infty} v_{\rm NS} = w ,
\label{eq:winf}
\end{equation}
where $v_2$ is the ejection velocity of the companion star
and $v_{\rm NS}$ is the ejection velocity of the newborn NS (both in the c.m. reference frame of the pre-SN binary).
As discussed in Section~\ref{sec:results}, in tight pre-SN systems ($r\rightarrow 0$) we have $v_{\rm 2,orb}^2 \gg v_{\rm im}^2$ which yields:
\begin{equation}
  \lim_{w \to \infty} v_2 \simeq v_{\rm 2,orb} = M_{\rm He}\,\sqrt{G/((M_{\rm He}+M_2)\,r)} .
\label{eq:v2proxy}
\end{equation}

In addition, we have for the case of a purely symmetric SN and neglecting the shell impact \citep[see also][]{tt98,gva09}: 
\begin{equation}
  \lim_{w, v_{\rm im} \to 0} v_2 = \sqrt{1-2\,M_{\rm NS}(M_{\rm He}+M_2)/M_{\rm He}^2}\; v_{\rm 2,orb}.
\label{eq:symm}
\end{equation}

\end{document}